\documentclass[aps,prl,preprint,superscriptaddress,showpacs]{revtex4-1}

\usepackage{graphicx}
\usepackage{dcolumn}
\usepackage{bm}

\usepackage{color}

\begin{document}



\title{Collimated Ultra-Bright Gamma-Rays from a PW-Laser-Driven Wire Wiggler}

\author{W.-M. Wang}
\affiliation{Beijing National Laboratory for Condensed Matter
Physics, Institute of Physics, CAS, Beijing 100190, China}
\affiliation{Beijing Advanced Innovation Center for Imaging
Technology and Key Laboratory of Terahertz Optoelectronics (MoE),
Department of Physics, Capital Normal University, Beijing 100048,
China}

\author{Z.-M. Sheng}
\affiliation{SUPA, Department of Physics, University of Strathclyde,
Glasgow G4 0NG, United Kingdom}\affiliation{Key Laboratory for Laser
Plasmas (MoE) and School of Physics and Astronomy, Shanghai Jiao
Tong University, Shanghai 200240, China} \affiliation{IFSA
Collaborative Innovation Center, Shanghai Jiao Tong University,
Shanghai 200240, China}\affiliation{Tsung-Dao Lee Institute,
Shanghai Jiao Tong University, Shanghai 200240, China}
\author{P. Gibbon}
\affiliation{Forschungzentrum J\"ulich GmbH, Institute for Advanced
Simulation, J\"ulich Supercomputing Centre, D-52425 J\"ulich,
Germany} \affiliation{Centre for Mathematical Plasma Astrophysics,
Katholieke Universiteit Leuven, 3000 Leuven, Belgium}
\author{L.-M. Chen}
\affiliation{Beijing National Laboratory for Condensed Matter
Physics, Institute of Physics, CAS, Beijing 100190, China}
\affiliation{IFSA Collaborative Innovation Center, Shanghai Jiao
Tong University, Shanghai 200240, China}

\author{Y.-T. Li}
\affiliation{Beijing National Laboratory for Condensed Matter
Physics, Institute of Physics, CAS, Beijing 100190, China}
\affiliation{IFSA Collaborative Innovation Center, Shanghai Jiao
Tong University, Shanghai 200240, China} \affiliation{School of
Physical Sciences, University of Chinese Academy of Sciences,
Beijing 100049, China}

\author{J. Zhang}
\affiliation{Key Laboratory for Laser Plasmas (MoE) and School of
Physics and Astronomy, Shanghai Jiao Tong University, Shanghai
200240, China} \affiliation{IFSA Collaborative Innovation Center,
Shanghai Jiao Tong University, Shanghai 200240, China}

\date{\today}

\begin{abstract}
It is shown by three-dimensional QED particle-in-cell simulation
that as a laser pulse of 2.5 PW and 20 fs propagates along a
sub-wavelength-wide solid wire, directional synchrotron
$\gamma-$rays along the wire surface can be efficiently generated.
With 8\% energy conversion from the pulse, the $\gamma-$rays
contains $10^{12}$ photons between 5 and 500 MeV within 10 fs
duration, corresponding to peak brilliance of $10^{27}$ photons
${\rm s^{-1}~ mrad^{-2}~ mm^{-2}}$ per 0.1\% bandwidth. The
brilliance and photon energy are respectively 2 and 3 orders of
magnitude higher than the highest values of synchrotron radiation
facilities. The radiation is attributed to the generation of nC, GeV
electron beams well guided along the wire surface and their wiggling
motion in strong electrostatic and magnetostatic fields induced at
the high-density-wire surface. In particular, these quasistatic
fields are so strong that QED effects already play a significant
role for the $\gamma-$ray radiation. With the laser power $P_0$
ranging from 0.5 PW to 5 PW available currently, this scheme can
robustly produce $\gamma-$rays peaked at $1^\circ$ with few-mrad
divergence and the photon energy and number roughly scales with
$P_0$ and $P_0^{3/2}$, respectively. Our scheme embraces both the
merits of high directionality comparable to those based upon laser
wakefield acceleration and high charge comparable to those based
upon laser-solid interaction.
\end{abstract}


\maketitle

Bright $\gamma-$rays with energy above MeV are highly demanded in
broad applications ranging from laboratory astrophysics
\cite{Bulanov}, emerging nuclear photonics \cite{np1,np2}, to
radiotherapy \cite{Cancer_radiotherapy,Radiotherapy}. These
applications can potentially benefit from $\gamma-$ray sources based
upon compact laser-driven electron acceleration. Via laser wakefield
acceleration (LWFA) \cite{Tajima,Esarey}, GeV electron beams
typically with duration of tens of fs, transverse size of
micrometers, and divergence of a few mrad are generated from gas
plasma. Through betatron radiation \cite{Rousse,Kneip,Cipiccia} or
Compton scattering
\cite{Malka_NP,Umstadter_PRL,Umstadter_NP,Umstadter_OL,Sarri,Khrennikov,LiuJS,Yan}
the beams are wiggled by electrostatic or/and laser fields and then
emit $\gamma-$rays basically with similar duration, size, divergence
to the beams. These cause high peak brilliance $10^{19}-10^{23}$
photons $\rm s^{-1}~ mrad^{-2}~ mm^{-2}$ per 0.1\% bandwidth (BW).
Mainly limited by wiggler field strengths, most $\gamma-$ray photons
are distributed in sub-MeV range. By increasing the scattering laser
strength \cite{Sarri,Yan} or frequency \cite{Umstadter_OL}, the
Compton photon energy can be enhanced to multi-MeV. However, both
the energy conversion efficiency from the pulse to the $\gamma-$rays
and the resulting photon number are not high, typically around
$10^{-6}$ \cite{Umstadter_PRL} and $10^6-10^8$ photons \cite
{Kneip,Cipiccia,Umstadter_NP,Sarri}, respectively, due to limited
charge in LWFA beams and wiggler strengths.

To overcome these limits and further enhance the photon energy to
the GeV range, we propose a scheme in which a currently-available PW
laser pulse \cite{Laser_4PW20fs,Laser_1PW_contrast,SULF_5PW}
propagates along a sub-wavelength wire, as shown in Figs. 1(a) and
1(b) (three-dimensional direct laser writing \cite{Wire_3D_print}
can provide wire arrays). Making use of the high density of the
wire, a directional GeV electron beam with tens of nC charge is
generated along its surface. Meanwhile, electrostatic and
magnetostatic fields induced at the surface are strong, which
intensively wiggles the beam electrons with significant QED
parameters. By QED synchrotron radiation from the GeV nC beam, 8\%
laser energy ($10^5$ higher than those based upon LWFA) is converted
to directional $\gamma-$rays containing $10^{12}$ photons with
energy up to GeV according to our 3D particle-in-cell (PIC)
simulations. The $\gamma-$rays, inheriting the laser duration and
wire width, have a high brilliance second only to X-ray free
electron lasers (XFEL), while the average photon energy of 20MeV is
3 orders of magnitude higher than XFEL, as shown in Fig. 1(f) and
Refs. \cite{European_ESRF,Shanghai_SSRF,Peak_SRF,XFEL}.

We show for the first time that the PW-laser-irradiated
sub-wavelength wire acts as a novel wiggler as well as an
accelerator of collimated electron beams of nC. Note that the wire
accelerator has been widely studied
\cite{Kodama,Tokita,Nakajima,YTLi,YMa}. Here, we show wiggling of
the beam electrons due to the electrostatic and magnetostatic fields
induced around the high-density wire surface, which are so high that
QED effects become significant. This is different from nonlinear
Compton scattering \cite{Ridgers,Brady,Wang-PRE17} or resonance
acceleration \cite{QiaoBin} in the QED regime, which is driven
directly by laser fields with powers above 10PW. Besides, in a
previous channel-like-target scheme \cite{Stark}, the wiggling
electrons are across the whole channel with the transverse size near
the laser spot diameter and therefore the generated photons have
emission angles of $40^\circ$. In our scheme the wiggling electrons
are restricted around the wire surface, which causes the photons
peaked at $1^\circ$ with few-mrad divergence. To our knowledge, our
scheme produces the $\gamma-$ray emission with the best directivity
so far based on laser-solid interaction.

We demonstrate our scheme [see Fig. 1(a)] through 3D PIC simulations
with the $\textsc{KLAPS}$ code \cite{KLAPS} including photon and
pair generation via QED processes \cite{Wang-PRE17,KLAPS-QED}. The
pulse propagates along the $+x$ direction with $y$-direction
polarization, wavelength $1\rm\mu m$ (period $\tau_0=3.3\rm fs$),
peak power 2.5PW, and FWHM duration 20fs. With an initial spot
radius $r_{ini}=6.12\rm\mu m$ and amplitude $a_{ini}=56$ normalized
by $m_ec\omega_0/e$ (intensity $4.3\times 10^{21}\rm~W cm^{-2}$),
the pulse is located at 5 Rayleigh lengths ($22.6\rm\mu m$) ahead of
the focusing plane. The spot radius at the focusing plane are
expected to be $r_{0}=1.2\rm\mu m$ with $a_0=285$ in the vacuum. An
aluminium wire of cuboid is taken with 50$\mu$m long in the x
direction and 0.6$\mu$m wide, which is placed 2.4$\mu$m behind the
pulse initial wavefront. It is assumed as fully-ionized plasma of
density $690n_c$ ($n_c=1.1\times10^{21}\rm~cm^{-3}$) with a
$0.2\rm\mu m$-exponential-scalelength preplasma. A moving window at
the light speed is taken with a simulation box $16\rm\mu m \times
24\rm\mu m \times 24\rm\mu m$ in $x\times y \times z$ directions. We
take the cell sizes in the three directions as $0.02\rm\mu m$, the
timestep as 0.033fs (adjustable timesteps for photon and pair
generation \cite{Wang-PRE17}), and 8 quasi-particles per cell.

Figures 1(c)-1(e) show the $\gamma-$rays emitted from the wire as
well as from a flat aluminium target with a large enough transverse
size of 24$\mu m$ for comparison.  With the wire, the $\gamma-$rays
have a sharp peak angle nearly along the wire surface [see Fig.
1(d)]. However, large divergence $\gamma-$rays are generated with
the flat target, as obtained in previous reports
\cite{Ridgers,Brady}. Although the energy conversion efficiencies
are similar in the two cases, the photon number in the peak angle is
one order of magnitude higher in the wire case. Figure 1(c) shows
that the $\gamma-$rays have FWHM duration about 10fs and a
transverse size near the wire width 0.6$\mu m$ because they are
generated around the wire surface. {\color{red} The brilliance is
$1.2\times10^{27}$, $8\times10^{26}$, and $1.5\times10^{26}$ photons
$\rm s^{-1}~ mrad^{-2}~ mm^{-2}$ 0.1\% BW at 5MeV, 20MeV, and
100MeV, respectively,} where the total $\gamma-$rays have
$1.75\times10^{10}$ photons in the angle $1^{\circ}$ with the
divergence of $\rm 3.49\times 3.49 ~mrad^2$ (we count the photon
number with an angle displacement of $0.2^{\circ}$). With the flat
target, the source size is increased to a few microns, determined by
the plasma area of laser hole boring \cite{Gibbon}. The increased
size and decreased photon number at the peak angle causes the peak
brilliance reduced by 3 orders of magnitude. Figure 1(e) shows the
photon spectrum distributed from 5MeV to 500MeV with average energy
about 20MeV in the wire case. Note that some beam electrons with
energy above 1GeV which can emit photons of 500MeV since the
electron QED parameters \cite{Erber,Elkina,Piazza} $\chi>0.2$ [see
Fig. 3]. With the flat target, both the photon energy and the number
in the higher-energy part are significantly reduced. This suggests
that the wire geometry is more favorable to bring larger $\chi$ for
higher photon energy.

We examine the wiggler fields in detail.  The fields composed of
electrostatic and magnetostatic components are perpendicular to
velocities of the beam electrons moving along the $+x$ direction.
First, the laser field strips a large number of electrons away from
the wire surface [see Fig. 2(b)], which induces electrostatic fields
$E_y^S$ [see Fig. 2(a)] and $E_z^S$ around the surfaces $y\simeq
\pm0.3\mu m$ and  $z\simeq \pm0.3\mu m$, respectively. In turn, the
laser field becomes hollow as observed in Fig. 1(b). Due to
radiation pressure, the hollow laser pulse together with the
electrostatic fields tends to confine electrons within the wire. To
compensate the beam-electron flux along the $+x$ direction, a return
current is formed around the wire surface [see Fig. 2(d)], which
induces magnetostatic fields $B_z^S$ [see Fig. 2(c)] around $y\simeq
\pm0.3\mu m$ and $B_y^S$ around $z\simeq \pm0.3\mu m$. According to
Figs. 2(a) and 2(c), $E_y^S$ and $B_z^S$ basically have similar
strengths and the same signs, positive at $y>0$ and negative at
$y<0$. For the electrons along $+x$ direction, the magnetic force is
opposite to the electric force, which can result in electron
wiggling along the $y$ direction with the force
$-e(E_y^S-v_{e,x}B_z^S)$. With $v_{e,x}\simeq1$, the wiggler field
around the surfaces $y\simeq \pm0.3\mu m$ can be written by
\begin{eqnarray}
F_y^{wiggler} \simeq E_y^S-B_z^S.
\end{eqnarray}
Note that contributions of laser electric and magnetic fields to
$F_y^{wiggler}$ (and resulting $\chi$ \cite{Bell}) are counteracted.
One can write $F_z^{wiggler} \simeq E_z^S+B_y^S$ around the surfaces
$z\simeq \pm0.3\mu m$.

Now we analyze if $F_y^{wiggler}$ can lead to effective wiggling
motion. Formation of the electrostatic and magnetostatic fields can
be described by $\partial E_y^S/\partial y + \partial E_z^S/\partial
z = 2\pi (n_i-n_e)$ and $\partial B_z^S/\partial y -
\partial B_y^S/\partial z=2\pi J_x$, where $E_x^S$, $B_x^S$, static
$J_y$ and $J_z$ are relatively weak as observed in our PIC
simulation. Here $n_i$ and $n_e$ are normalized by $n_c$, $J_x$ by
$ecn_c$, and fields by $m_ec\omega_0/e$. According to our PIC
simulation, we find that $E_z^S$, $B_y^S$, $\partial E_z^S/\partial
z$, and $\partial B_y^S/\partial z$ are roughly constant at the
surface with a given $z$ since the wire width are much smaller than
the laser spot diameter [one can similarly see in Figs. 2(a) and
2(c) that $E_y^S$, $B_z^S$, $\partial E_y^S/\partial y$ and
$\partial B_z^S/\partial y$ are roughly constant at the surface with
a given $y$]. Then, $\partial E_y^S/\partial y \simeq 2\pi
(n_i-n_e-\alpha_1)$ and $\partial B_z^S/\partial y=2\pi
(J_x-\alpha_2)$ at a given $z_0$, where $\partial E_z^S/\partial
z|_{z_0} \simeq 2\pi \alpha_1$ and $-\partial B_y^S/\partial
z|_{z_0}=2\pi \alpha_2$. One can obtain:
\begin{eqnarray}
\partial F_y^{wiggler}/\partial y \simeq 2\pi
(n_i-n_e-J_x-\alpha_1+\alpha_2)=2\pi \rho^{eff}.
\end{eqnarray}
According to this equation, one can understand Figs. 2(e) and 2(f),
where we simply take $\alpha_1=40$ and $\alpha_2=30$ to satisfy
neutrality at $y=0$. Note that basically $|\alpha_1-\alpha_2|$ is
far smaller than $|n_i-n_e|$ and $|J_x|$, so that the effective
charge density $\rho^{eff}$ is mainly determined by $n_i-n_e-J_x$.
Around the wire center, $\rho^{eff}\simeq0$; Increasing $|y|$,
electrons are piled up by laser radiation pressure with $n_e>n_i$
and return currents are mainly located this region with $J_x>0$, and
consequently $\rho^{eff}<0$; Further increasing $|y|$ and close to
the surface, wire electrons are stripped with $n_e\sim 0$, there are
well-guided beams in the ion channel with $J_x<0$, and thus
$\rho^{eff}\simeq n_i-J_x>0$ [see Fig. 2(f)].

Such $\rho^{eff}$ generates effective wiggler fields $F_y^{wiggler}$
shown in Fig. 2(e). There are two zero-field points close to the
surfaces $y\simeq \pm0.3\mu m$, respectively. Around these points
the fields are bipolar, which naturally causes electron wiggler.
Note that the peak field strength inside the wire is higher than
that outside, which prevents the beam electrons from crossing the
wire center and keeps them wiggling at one side of the wire [see
Fig. 3(a)]. One can also see in Fig. 2(e) that change of
$F_y^{wiggler}$ with $y$ is sharp at the zero-field points due to
large $\rho^{eff}\simeq n_i$. This causes small spatial displacement
of the electron wiggling motion and small angles of photon emission
[see Figs. 3(a) and 3(c)].

The trajectory and energy evolution for a test electron located
around the wire surface $y\simeq -0.3\mu m$ are plotted in Figs.
3(a) and 3(c). One can see in Fig. 3(a) that the field $E_y-B_z$
experienced by the electron varies with the transverse position $y$
but not with $x$ since it moves along with the laser at
$v_{e,x}\simeq 1$. This suggests that its wiggling motion is driven
by the static fields rather than the laser fields. As the pulse
moves to the focusing plane $x=25\mu m$, the electron energy
$\varepsilon$ grows gradually to $>$1GeV with increasing QED
parameter $\chi$ and decreasing emission angles $\theta$. Around the
focusing plane, the strongest emission arises with the largest
$\chi\simeq 0.2$ accompanied with the smallest $\theta \simeq
1^{\circ}$ and therefore the $\gamma-$rays have the angle peak
around $1^{\circ}$ [see Figs. 1(d) and 4(a)]. At later, both
$\varepsilon$ and $\chi$ decrease while $\theta$ increases. This is
why we take the laser focusing plane a few Rayleigh lengths behind
the wire fore-end, allowing a distance to accelerate and generate
well-guided GeV beam before the highest laser intensity and
resulting the largest $\chi$. The QED parameter $\chi= \gamma_e
\sqrt{(\mathbf{E}+\mathbf{v_e}\times
\mathbf{B})^2-(\mathbf{v_e}\cdot \mathbf{E})^2}/E_{Sch}$
\cite{Erber,Elkina,Piazza} of an electron with $v_{e,x}\simeq 1$ can
be simplified as
\begin{eqnarray}\label{wiggler_field}
\chi \simeq \gamma_e |E_y^S-B_z^S|/E_{Sch},
\end{eqnarray}
for the wiggler along the y direction, where
$E_{Sch}=1.32\times10^{18}V/m$ is the Schwinger strength
\cite{Schwinger1,Schwinger2}, and $\gamma_e$ and $\mathbf{v_e}$ are
electron relativistic factor and velocity normalized by $c$.
According to Eq. (3) with $|E_y^S-B_z^S|\simeq 50$, $\gamma_e\simeq
1957$ read from Figs. 3(a) and 3(c), one can calculate $\chi=0.23$
in agreement with Fig. 3(c).

Figure 4 indicates that our scheme is robust. Similar photon angular
distributions are achieved when the power is ranging from 0.5PW to
5PW available currently \cite{Laser_4PW20fs,SULF_5PW} and the width
from $0.5\mu m$ and $0.8\mu m$ (even with similar conversion
efficiencies). In particular, even at 0.5PW the $\gamma-$ray
brilliance can reach {\color{red}$1.2\times10^{26}$ photons $\rm
s^{-1}~ mrad^{-2}~ mm^{-2}$ 0.1\% BW at 6MeV}. When the width is too
small, e.g., $0.1\mu m$, the wire is completely destructed by the
laser fields and electrons move like in the vacuum. Hence, the
$\gamma-$rays have high divergence and low conversion efficiency.
When increasing the width to $0.3\mu m$, the wire structure can be
kept before the pulse approaches its focusing plane. Then, electrons
are first wiggled around the wire surface and later cross the wire
center with large angles when strongest radiation occurs due to
$\varepsilon$ and $\chi$ at the maximums [see Figs. 3(b) and 3(d)].
This causes the $\gamma-$rays peaked at a larger angle than the
$0.6\mu m$ wire case [see Fig. 4(a)]. These can be seen more clearly
in Figs. 3(e)-3(h) which show spatial, angular, energy distributions
of electrons. In the $0.6\mu m$ case [Figs. 3(e) and 3(g)], the
higher-energy electrons are distributed around the surface and
peaked at $1^{\circ}$, which have nC charge. They are wiggled on one
side of the surface and then strongly emit $\gamma-$rays around
$1^{\circ}$. In the $0.3\mu m$ case [Figs. 3(f) and 3(h)], however,
the electrons are peaked at the wire center and around $10^{\circ}$,
where the pulse is absorbed more strongly since it can enter the
wire interior, rather than being stopped by the surface in the
$0.6\mu m$ wire. Thus, the conversion efficiency appears highest
around $0.3\mu m$ [see Fig. 4(c)]. For the same reason, the
efficiency decreases with the increasing wire width.

To further understand Figs. 4(d) and 4(e), we analyze the photon
energy and number scaling with the laser power. The electron beam
energy can be given by $\langle \gamma_e \rangle \simeq 3.13
a_0\exp(-\lambda_0^2/16r_0^2)$ according to Ref. \cite{YMa}, which
predicts the value 437MeV close to the peak energy 650MeV shown in
Fig. 3(g). Then, Eq. (3) can be rewritten by $\langle \chi \rangle
\simeq 3.13 a_0\exp(-\lambda_0^2/16r_0^2) |F_y^{wiggler}|/E_{Sch}$.
In our case with the peak intensity around $10^{23}\rm~W cm^{-2}$
and the wire width below $\lambda_0$, the electrons on the wire
surface are quickly stripped and therefore, the static field
strength or $|F_y^{wiggler}|$ depends strongly upon the wire charge
density and weakly upon the laser intensity, as observed in our
simulations and Eq. (2). When the wire parameter is fixed and the
laser power $P_0$ is adopted within 0.5 to 5PW, one can roughly take
$|F_y^{wiggler}|$ as a value about 50 according to our simulations
and then $\langle \chi \rangle \simeq 0.00037 a_0$. To obtain photon
data, one can use the theory of synchrotron radiation
\cite{Bell,Ridgers}, which is general when the acceleration field of
an electron is given in its rest frame, i.e., $\chi$. The emitted
photons have an average energy
\begin{eqnarray}
\langle \varepsilon_{ph} \rangle = 0.44\langle \chi \rangle \langle
\gamma_e \rangle m_ec^2 \simeq 0.000245 a_0^2 ~ \rm [MeV]
\end{eqnarray}
and the photon generation rate per electron is $1.4\times 10^{13}
\langle \gamma_e \rangle \simeq 4.2\times 10^{13} a_0$. With $P_0=$
5, 2.5, 1, 0.5PW, $\langle \varepsilon_{ph} \rangle$ is calculated
as 40, 20, 8, 4MeV, respectively, which reasonably agrees with our
simulation results: 31, 20, 13, 6MeV. To obtain the photon number,
we count the number $N_e$ of electrons above 10MeV in our
simulations and find a rough scaling $N{e}\propto a_0^2$. We assume
that beam electrons have nearly the same efficient radiation time
with $P_0$ ranging from 0.5PW to 5PW, since the pulse spot size is
much larger than the wire width and therefore the wire slightly
affects the evolution of the pulses with different $P_0$. Then, the
photon number follows
\begin{eqnarray}
N_{ph}\propto a_0^3,
\end{eqnarray}
which agrees with our simulation results: $2.8\times 10^{12}$,
$1.24\times 10^{12}$, $3.6\times 10^{11}$, and $1.6\times 10^{11}$
photons with 5, 2.5, 1, and 0.5 PW, respectively. From Eqs. (4) and
(5), one can obtain the conversion efficiency $\eta \propto a_0^3$,
in reasonable agreement with the results shown in Fig. 4(d).

In summary, we have shown that a PW-laser-irradiated sub-wavelength
solid wire acts as a novel wiggler and an accelerator of nC, GeV,
high-directivity electron beams. The wiggler is driven by
electrostatic and magnetostatic fields around the wire surface,
rather than directly by the laser fields. Due to high density of the
wire, the quasistatic fields are so high that the wiggling electrons
have $\chi>0.1$. With the synchrotron radiation in the QED regime,
ultra-bright, tens-of-MeV, few-mrad-divergence $\gamma-$rays peaked
at $1^\circ$ can be efficiently generated with $P_0$ between 0.5PW
and 5PW. The average photon energy scales linearly with $P_0$ and
the photon number and conversion efficiency with $P_0^{3/2}$. In our
scheme, the laser focusing plane should be behind the wire fore-end,
allowing a distance to generate well-guided GeV beams before
achieving the largest $\chi$.

\begin{acknowledgments}
This work was supported by Science Challenge Project of China (Grant
No. TZ2016005), the National Basic Research Program of China (Grants
No. 2013CBA01500), the National Natural Science Foundation of China
(Grants No. 11775302, No. 11375261, No.11421064, No. 11374210, No.
113111048, and No.11520101003), and the Strategic Priority Research
Program of the Chinese Academy of Sciences (Grants No. XDB16010200
and No.XDB07030300). Z.M.S. acknowledges the support of a Leverhulme
Trust Research Grant and EPSRC (UK) Grant No. EP/N028694/1, and EC's
H2020 EuPRAXIA (Grant No. 653782).
\end{acknowledgments}

\newpage

\begin{figure}[htbp]
\includegraphics[width=6.5in]{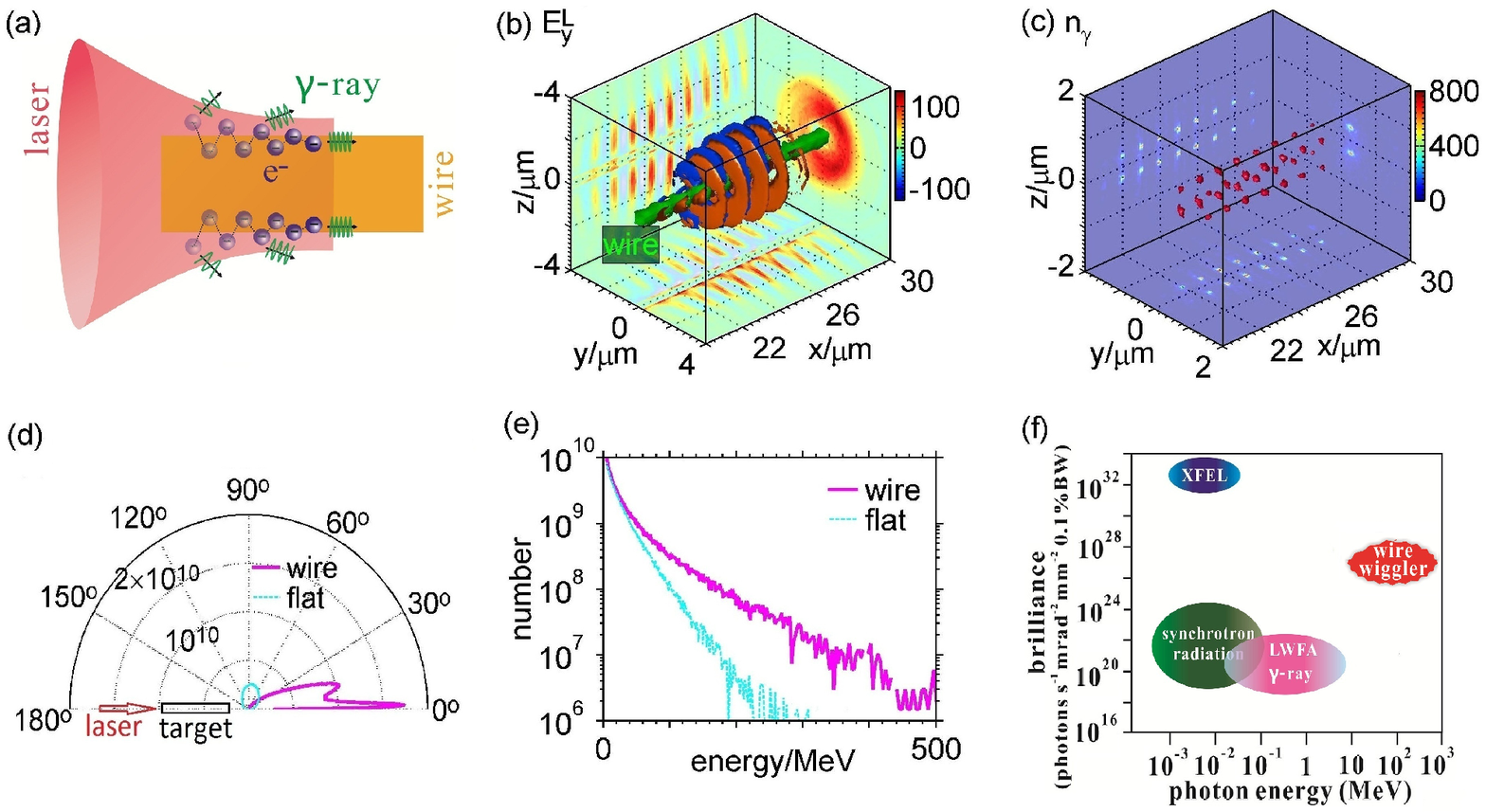}
\caption{\label{fig:epsart} (a) Schematic: as a laser propagates
along a subwavelength wire and approaches its focusing plane,
electrons along the wire surface are gradually accelerated with
reduced divergent angles, meanwhile, the electrons are wiggled
perpendicularly to the surface, which causes $\gamma-$rays emitted
with increased photon energies and decreased divergent angles.
Three-dimensional isosurfaces of (b) the laser field (units of
$mc\omega_0/e$) and (c) $\gamma-$ray photon density (units of $n_c$)
at the time of $30\tau_0$ as well as the slices at the planes with
respective peak values, where a $0.6\mu m$-wide wire is taken. (d)
Angular distributions and (e) spectra of $\gamma-$rays emitted from
the wire and a flat target, respectively. (f) Photon energy and
brilliance of $\gamma-$rays generated from synchrotron radiation
facilities, XFEL \cite{XFEL}, betatron radiation and Compton
scattering based on LWFA, and our scheme.}
\end{figure}

\begin{figure}[htbp]
\includegraphics[width=6.5in]{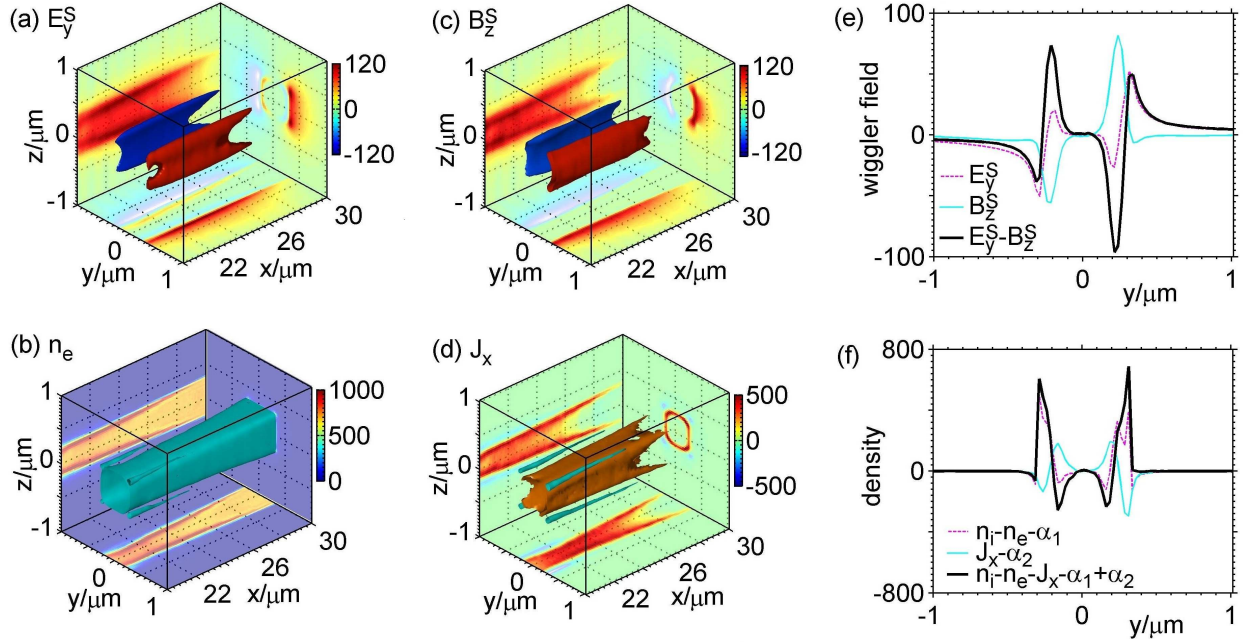}
\caption{\label{fig:epsart} Three-dimensional isosurfaces of (a)
electrostatic and (c) magnetostatic fields (units of
$mc\omega_0/e$), (b) electron density (units of $n_c$), and (d)
current density (units of $ecn_c$) at the time of $30\tau_0$ as well
as the slices at the planes with respective peak values, where they
are obtained by temporally averaging $E_y$, $B_z$, $n_e$, and $J_x$,
respectively, over one laser cycle. The corresponding
one-dimensional distributions of these fields and densities at
$x=21\mu m$ and $z=0.26\mu m$ are shown in (e) and (f).}
\end{figure}

\begin{figure}[htbp]
\includegraphics[width=6.5in]{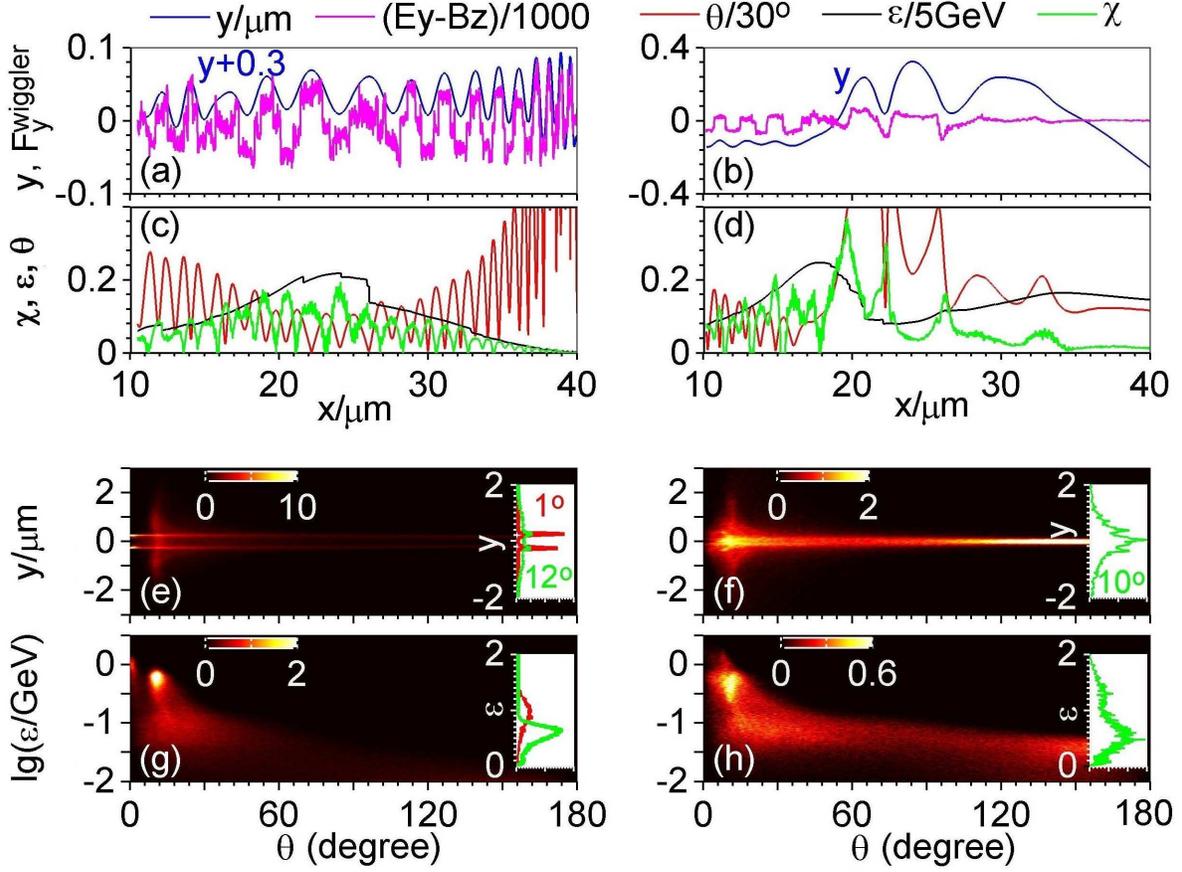}
\caption{\label{fig:epsart} Evolution for a test electron from the
$0.6\mu m$ [(a), (c)] and $0.3\mu m$ wires [(b), (d)], respectively,
is shown of the transverse position $y$ (units of $\mu m$),
divergence angle $\theta$ (units of $30^o$), energy $\varepsilon$
(units of 5GeV), QED parameter $\chi$, and $E_y-B_z$ (units of
$1000m_ec\omega_0/e$), where we plot $y+0.3$ in (a) since the
electron wiggles around $-0.3~\mu m$. (e)-(h) Number (units of
$10^8$) of electrons $>$10MeV as a function of ($\theta$, $y$,
$\varepsilon$) at $30\tau_0$, where insets in each plot show number
distributions at given angles. The left and right columns correspond
to $0.6\mu m$ and $0.3\mu m$ wires, respectively.}
\end{figure}

\begin{figure}[htbp]
\includegraphics[width=6.5in]{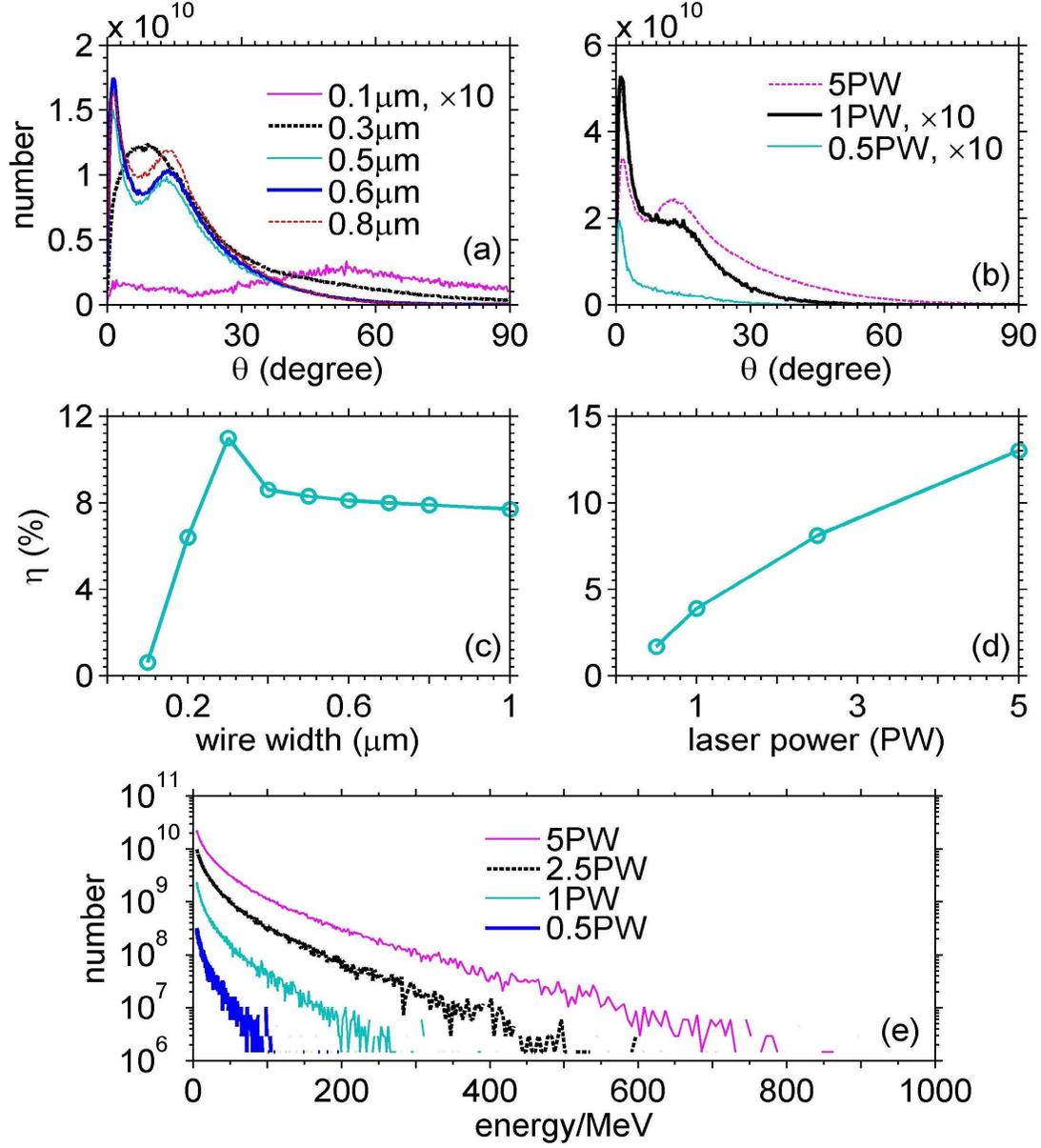}
\caption{\label{fig:epsart} Angular distributions of $\gamma-$rays
with different wire widths (a) and laser powers (b), where ``$\times
10$'' in the legend means the number multiplied by 10. $\gamma-$ray
conversion efficiency versus (c) wire widths and (d) laser powers.
(e) $\gamma-$ray spectra at $50\tau_0$ under different powers. In
(a) and (c), the power is fixed at 2.5PW. In (b), (d), and (e), the
wire width is fixed at $0.6\mu m$.}
\end{figure}

\end{document}